# Quantum Transport Simulation of Sub-1-nm Gate Length Monolayer MoS$_2$ Transistors

Ying Li,[1,†] Yang Shen,[2,†] Linqiang Xu,[1] Shiqi Liu,[3] Yang Chen,[1] Qiuhui Li,[1] Zongmeng Yang,[1] Xiaotian Sun,[4,*] He Tian,[2,*] and Jing Lu,[1,5,6,7,8,9,*]

[1] State Key Laboratory for Mesoscopic Physics and Department of Physics, Peking University, Beijing 100871, P. R. China

[2] Institute of Microelectronics and Beijing Research Center for Information Science and Technology (BNRist), Tsinghua University, Beijing 100084, China

[3] Key Laboratory of Spintronics Materials, Devices and Systems of Zhejiang Province, Hangzhou 311305, China

[4] College of Chemistry and Chemical Engineering, and Henan Key Laboratory of Function-Oriented Porous Materials, Luoyang Normal University, Luoyang 471934, P. R. China

[5] Collaborative Innovation Center of Quantum Matter, Beijing 100871, P. R. China

[6] Beijing Key Laboratory of Quantum Devices, Peking University, Beijing 100871, P. R. China

[7] Beijing Key Laboratory for Magnetoelectric Materials and Devices (BKL-MEMD), Peking University, Beijing 100871, P. R. China

[8] Peking University Yangtze Delta Institute of Optoelectronics, Nantong 226010, P. R. China

[9] Key Laboratory for the Physics and Chemistry of Nanodevices, Peking University, Beijing 100871, P. R. China

† These authors contributed equally to this work.

*Corresponding Authors: jinglu@pku.edu.cn (Jing Lu); tianhe88@tsinghua.edu.cn (He Tian); sxt@lynu.edu.cn (Xiaotian Sun)

# Abstract

Sub-1-nm gate length MoS$_2$ transistors have been experimentally fabricated, but their device performance limit remains elusive. Herein, we explore the performance limits of the sub-1-nm gate length monolayer (ML) MoS$_2$ transistors through *ab initio* quantum transport simulations. Our simulation results demonstrate that, through appropriate doping and dielectric engineering, the sub-1-nm devices can meet the requirement of extended "ITRS (International Technology Roadmap for Semiconductors) $L_g$=0.34 nm". Following device optimization, we achieve impressive maximum on-state current densities of 409 µA/µm for *n*-type and 800 µA/µm for *p*-type high-performance (HP) devices, while *n*-type and *p*-type low-power (LP) devices exhibit maximum on-state current densities of 75 µA/µm and 187 µA/µm, respectively. We employed the Wentzel-Kramer-Brillouin (WKB) approximation to explain the physical mechanisms of underlap and spacer region optimization on transistor performance. The underlap and spacer regions primarily influence the transport properties of sub-1-nm transistors by respectively altering the width and body factor of the potential barriers. Compared to ML MoS$_2$ transistors with a 1 nm gate length, our sub-1-nm gate length HP and LP ML MoS$_2$ transistors exhibit lower energy-delay products. Hence the sub-1-nm gate length transistors have immense potential for driving the next generation of electronics.

**Keywords:** MoS$_2$, Sub-1-nm gate length, transistor, scaling, quantum transport simulation

# 1 Introduction

Semiconductor technology continues to advance, with silicon remaining the predominant material [1,2]. The 3-nanometer (nm) node technology has already entered mass production, and the 2-nm node technology is progressing as scheduled [3]. Furthermore, it is anticipated that 1-nm technology will be introduced ahead of schedule by IMEC in 2028 [4]. However, as feature sizes continue to shrink, traditional silicon-based field-effect transistors (FETs) struggle to maintain reliable and robust performance when the physical gate length falls below 5 nm, mainly due to short-channel effects (SCEs) [5,6]. Two-dimensional (2D) materials present a novel approach to further reduce the physical dimensions of FETs, leveraging their atomic-scale thickness for favorable electrostatics and their flat, dangling-bond-free surface for high carrier mobilities [7-9].

Considerable efforts have been devoted to realizing 2D FETs with extremely small dimensions [10,11]. Among these efforts, 2D $MoS_2$ has emerged as the most extensively researched material, both theoretically and experimentally [12-15]. Notably, Desai *et al.* successfully fabricated bilayer $MoS_2$ FETs with a gate length of 1 nm, employing a single-walled carbon nanotube (SWCNT) as the gate electrode [16]. The resulting 1-nm-gate device exhibited an excellent subthreshold swing (SS) of 65 mV/dec but featured a saturation current of approximately only 10 µA/µm. More recently, Ren and Tian *et al.* experimentally realized a sub-1-nm gate length monolayer (ML) $MoS_2$ transistor by utilizing the sidewall of ML graphene as the gate electrode [17]. The lowest SS and highest saturation current among these sub-1-nm-gate devices are 117 mV/dec and ~$10^{-1}$ µA/µm, respectively.

These saturation currents fall well below the on-state current standards in ITRS (International Technology Roadmap for Semiconductors) [18] and IRDS (International Roadmap for Devices and Systems) [19], rendering them unsuitable for both high-performance (HP) and low-power (LP) applications. The above-mentioned 1-nm and sub-1-nm gate length transistors primarily highlight their gate lengths; however, their channel lengths remain around 500 nm, and no optimization has been carried out for their contact electrodes. Consequently, they are far from reaching the performance limits of sub-1-nm gate length 2D $MoS_2$ transistors. Recent studies reported ohmic-contacted three-layer InSe [20] and three-layer $MoS_2$ [21] FETs with channel lengths of 10 nm, showcasing impressive ballistic ratios of 83% and 80%, respectively. Their on-state currents can reach as high as 1200 and 1220 μA/μm, respectively. These works provided technical routes for optimizing sub-1-nm gate length devices in experiments. Therefore, it is imperative to investigate the performance limits of ML MoS$_2$ FETs, by synergizing Ohmic contacts, ballistic transport, and sub-1-nm gate lengths.

In this study, we utilized a first-principles quantum transport method to simulate the performance limits of ML MoS$_2$ FETs with an ultra-small gate length of 0.34 nm (equivalent to the thickness of an ML graphene). Through optimization of the underlap length and the dielectric constant of the spacer region, we achieved further enhancements in device performance. Additionally, we investigated the physical mechanisms through which the underlap and spacer regions influence the transport properties of sub-1-nm devices, employing the Wentzel-Kramer-Brillouin (WKB) approximation. We found that the underlap and spacer regions predominantly affect the transport properties of sub-1-nm transistors by respectively altering the width and body factor of the potential barriers. Our simulations predicted that to meet the extended ITRS standards for *n*-type HP and LP devices, minimum underlap lengths of 20.7 and 23.2 Å, respectively, are required. The on-state currents of the *n*-type HP and LP devices can reach 409 and 75 μA/μm, respectively. For the *p*-type devices, the minimum underlap lengths needed to comply with the extended ITRS standards for HP and LP applications are 10.7 and 20.7 Å, respectively, and the maximum on-state currents for HP and LP applications are 800 and 187 μA/μm, respectively. These findings underscore the promising potential of sub-1-nm gate FETs in further reducing the physical dimensions of transistors. Compared to the 1-nm-gate ML MoS$_2$ FETs with the same doping type, our sub-1-nm gate ML MoS$_2$ FETs exhibited lower energy-delay products for both HP and LP applications. This suggests that sub-1-nm-gate FETs can strike a better balance between energy consumption and switching speed.

## 2 Method

In **Figure 1 (a)**, we illustrate the application of a two-probe model, which includes a central region, a left electrode, and a right electrode. The electrodes are considered as semi-infinite bulks, while the central region is treated as an open system. The Hamiltonians for both the electrodes and the central region are constructed using density functional theory (DFT) [22,23]. Subsequently, we employ the non-equilibrium Green's function (NEGF) formalism [24,25] to compute the non-equilibrium density matrix of the central region. The density matrix and Green's function of the central region are solved in a self-consistent manner. This DFT+NEGF formalism [24-26] has been integrated into the computational software Quantum Atomistix ToolKit 2020 [27-29].

In DFT, the Hamiltonians of the left/right electrodes, denoted as $H_{L/R}$, and the central region, denoted as $H$, are functionals of the electron densities, represented as follows:

$$\begin{aligned} H(n) &= -\frac{1}{2}\nabla^2 + V_{\text{ext}} + V_{\text{H}}(n) + V_{\text{xc}}(n) \\ H_{L/R}(n_{L/R}) &= -\frac{1}{2}\nabla^2 + V_{\text{ext}} + V_{\text{H}}(n_{L/R}) + V_{\text{xc}}(n_{L/R}) \end{aligned} \quad (1)$$

Here, $n$ and $n_{L/R}$ refer to the electron densities of the left/right electrode and the central region, respectively. The terms $V_{\text{ext}}$, $V_{\text{H}}$, and $V_{\text{xc}}$ correspond to the external potential, the Hartree potential, and the exchange-correlation potential, respectively. The Hartree potential in equation (1) is obtained by solving the Poisson equation for the electron density $n(\mathbf{r})$ and $n_{L/R}(\mathbf{r})$:

$$\nabla^2 V_{\text{H}} = -\frac{e^2}{4\pi\varepsilon_r\varepsilon_0} n(\mathbf{r}) \quad (2)$$

where $e$, $\varepsilon_r$, and $\varepsilon_0$ are the elementary charge, relative permittivity, and vacuum's permittivity, respectively. The electrode regions and the central region all have periodic boundary conditions and Neumann boundary conditions in the x and y directions, respectively. In the z direction, the left and right electrodes are considered to be equilibrium bulk materials with periodic boundary conditions, while the central region adopts Dirichlet boundary conditions [30].

The Green's function $G_{L/R}$ for the left/right electrode is given by:

$$G_{L/R} = (EI - H_{L/R} + i\eta)^{-1} \quad (3)$$

In this equation, $E$, $I$, $i$, and $\eta$ are energy, the identity matrix, the imaginary unit, and a positive infinitesimal, respectively. The impact of the electrodes on the central region is manifested through the self-energy terms:

$$\Sigma_{L/R} = \tau_{L/R} G_{L/R} \tau_{L/R}^\dagger \quad (4)$$

Here, $\Sigma_{L/R}$ represents the self-energy, and $\tau_{L/R}$ stands for the coupling between the left/right electrode and the central region. Subsequently, the Green's function $G$ for the central region is calculated as:

$$G = (EI - H - \Sigma_\text{L} - \Sigma_\text{R} + i\eta)^{-1} \tag{5}$$

The density matrix $\rho$ for the central region can be computed from the Green's function as follows:

$$\rho = \int \left(\frac{dE}{2\pi}\right) G[f(E - \mu_\text{L})\Gamma_\text{L} + f(E - \mu_\text{R})\Gamma_\text{R}]G^\dagger$$
$$\Gamma_\text{L/R} = i(\Sigma_\text{L/R} - \Sigma_\text{L/R}^\dagger) \tag{6}$$

where $\Gamma_\text{L/R}$, $f$ and $\mu_\text{L/R}$ are the imaginary part of the self-energy term (broadening function), the Fermi distribution function, and the chemical potential of the left/right electrode, respectively.

After solving the Green's function $G$ and density matrix $\rho$ of the central region self-consistently, the current $I$ can be determined using a formula similar to the Landauer-Büttiker formula [31]:

$$I = \frac{-2e}{h}\int_{-\infty}^{\infty} T(E)[f(E - \mu_\text{R}) - f(E - \mu_\text{L})]\, dE \tag{7}$$

Here, $h$ represents Planck's constant, and $T(E)$ stands for the transmission function. In practice, electrons with energy $E$ may correspond to various wavevector $k$ directions. Because the wavevector in the transport direction changes at the interface, we use the in-plane wavevector $k_\parallel$ perpendicular to the transport direction as a representative parameter for the wavevector. For each wavevector $k_\parallel$, the transmission coefficient $T(k_\parallel, E)$ can be obtained from the corresponding part of the Green's function $G(k_\parallel, E)$ and broadening function $\Gamma_\text{R}(k_\parallel, E)$:

$$T(k_\parallel, E) = \text{Tr}[\Gamma_\text{L}(k_\parallel, E)G(k_\parallel, E)\Gamma_\text{R}(k_\parallel, E)G(k_\parallel, E)^\dagger] \tag{8}$$

The transmission function $T(E)$ at energy $E$ is the sum of contributions from all the wavevectors $k_\parallel$ in the first Brillouin zone:

$$T(E) = \sum_{k_\parallel} w_{k_\parallel} T(k_\parallel, E) \tag{9}$$

Here, $w_{k_\parallel}$ represents the weight associated with $k_\parallel$.

We employed the generalized gradient approximation (GGA) Perdew-Burke-Ernzerhof (PBE) exchange-correlation functional [32] in our calculations. For structural optimization and electronic structure calculations of the ML MoS$_2$, a 20 Å vacuum layer was included. We utilized a plane wave (PW) basis set and the projector augmented wave method (PAW) [33], and the energy cutoff was set to 544 eV. Regarding k-point sampling in reciprocal space, we used a Monkhorst-Pack (MP) grid [34] of 11×11×1 during the optimization process and 17×17×1 for the electronic structure calculations. The convergence criteria for structural optimization were an energy difference below 10$^{-5}$ eV and a force on each atom below 10$^{-2}$ eV·Å$^{-1}$. For electronic structure calculations, the convergence criteria were an energy difference below 10$^{-6}$ eV. In the transport calculations, we employed norm-conserving (NC) PseudoDojo pseudopotentials [35] and the linear combination of atomic orbitals (LCAO) basis set. The real-space cutoff energy was set to 80 Hartree, and the electron temperature was 300 K. During the self-consistent process, the electrodes were treated with an MP grid of 17×1×173 while the central region used an MP grid of 17×1×1. For the calculation of the transmission coefficient, we use a denser MP grid of 32×1×1 to ensure convergence.

In general, GGA functionals, which are based on the one-electron approximation, tend to underestimate the band gap of free-standing intrinsic semiconductors. However, in the context of 2D FETs, the presence of the dielectric environment and carriers from electrodes strongly screens the electron-electron interactions [36]. This phenomenon makes the GGA PBE functional applicable, as previously reported [36-38]. For instance, in GW calculations [39,40], the quasi-particle bandgap of free-standing ML MoS$_2$ is found to be 2.8 eV, whereas, within the device environment, this bandgap is reduced to 1.9 eV [36]. This value is in proximity to the bandgap of 1.7 eV calculated by the GGA PBE functional [41]. Experimental and simulated data for transfer characteristics, on-state currents, delay times, and energy consumptions of sub-10-nm 2D carbon nanotube FETs [42-44], MoS$_2$ FETs [12,21,45], and InSe FETs [20,46] exhibit excellent agreement, affirming the reliability of the DFT (GGA PBE) + NEGF method.

## 3 Results and Discussion

### 3.1 Evaluation Parameters for Sub-1-nm Transistors

Due to the challenges associated with shrinking the physical dimensions of transistors, both the ITRS and IRDS have gradually relaxed their requirements for scaling down the physical gate length, as illustrated in **Figure S1** (supporting information [47]). In the most recent IRDS 2022 [19], no performance standards are outlined for devices with gate lengths below 1 nm, and the shortest gate length specified in this standard is 12 nm. In contrast, the ITRS 2013 [18] provides performance standards for devices with a minimum gate length of 5 nm. As a consequence, by extrapolating the performance standards from the ITRS 2013 guidelines for sub-10-nm gate length FETs, we obtained a performance standard reference for devices with a gate length of 0.34 nm, as depicted in **Figure S2** and summarized in **Table S1** (supporting information [47]). We refer to this standard as the "extended ITRS standard" and utilize it to evaluate the performance of our devices.

We determine the off-state voltage $V_{\text{off}}$ as the voltage on the transfer characteristic curve where the current equals the off-state current $I_{\text{off}}$ in the extended ITRS. Then, we calculate and obtain the on-state voltage $V_{\text{on}}$ as $V_{\text{on}} = V_{\text{off}} \pm V_{\text{dd}}$ for the *n*-type and *p*-type doping, respectively. $V_{\text{dd}}$ represents the driven voltage. The current at $V_{\text{on}}$ is the on-state current $I_{\text{on}}$.

**3.2 Device Structure and Doping Concentration Optimization**

We utilized intrinsic ML MoS$_2$ as the channel material, and its lattice constant was optimized to a=b=3.16 Å, which closely aligns with the previously reported value of 3.18 Å [48]. As displayed in **Figure S3** (supporting information [47]), we computed the band structure and projected density of states (PDOS) for ML MoS$_2$, obtaining a bandgap of 1.75 eV, an electron effective mass of 0.47 m$_0$, and a hole effective mass of 0.59 m$_0$. These values are consistent with the previously reported bandgap of 1.7 eV, an electron effective mass of 0.43 m$_0$, and a hole effective mass of 0.55 m$_0$ found in the literature [41,49].

To explore the performance limits of sub-1-nm gate length ML MoS$_2$ FETs, we used doped ML MoS$_2$ as the source and drain electrodes to directly inject charge carriers. As illustrated in **Figure 1 (a)**, we adopted a double-gate (DG) structure with a fixed gate length ($L_g$) of 0.34 nm (equivalent to the thickness of an ML graphene wall, representing the shortest achievable gate length experimentally [17]). In this configuration, we introduced underlap (UL) regions to mitigate drain-induced barrier lowering (DIBL) [50,51] and spacer regions (spacer) to enhance gate control [52,53]. We optimized the doping concentration, the length of the underlap region ($L_{\text{UL}}$), and the dielectric constant of the spacer regions ($\varepsilon_s$) to maximize the performance of

sub-1-nm gate length DG ML MoS$_2$ FETs. In the simulation, the central region has a total length of 76.6 Å in the z-direction, including the lengths of two electrode extension regions, two underlap regions, and the gate length. We adjusted the lengths of the doped electrode extension regions and the UL regions to modify the channel length, with the total channel length given as $L_{\text{ch}} = L_{\text{g}} + 2 L_{\text{UL}}$. The equivalent oxide thickness (EOT) of the UL region was 0.26 nm, while its physical thickness was 1.33 nm when we selected the dielectric constant of the underlap regions to be 20 (matching the dielectric constant of HfO$_2$). The upper and lower spacer regions covered the entire channel (comprising the undoped intrinsic ML MoS$_2$) in the z-direction, with each spacer region having a height of 27 Å in the y-direction.

To effectively enhance the saturation current of the device, a higher doping concentration is preferred. However, this can also make it more challenging to turn off the device. Therefore, it is crucial to select an appropriate doping concentration to optimize the on-state current. We kept $L_{\text{g}}$, $L_{\text{UL}}$, and $\varepsilon_{\text{s}}$ fixed at 0.34 nm, 23.2 Å, and 20, respectively, and conducted simulations of transfer characteristic curves for devices with varying doping concentrations in the electrodes, as displayed in **Figure 1 (b)**. Due to the limited range of stable doping concentrations in 2D materials [54], we explored doping concentrations within the range of $1 \times 10^{12} \sim 4 \times 10^{13}$ cm$^{-2}$. The on-state currents of both HP and LP devices, obtained from the transfer characteristic curves at different doping concentrations, are displayed in **Figure 1 (c)**. As depicted in the figure, when the doping concentration is $2 \times 10^{13}$ cm$^{-2}$, the on-state currents for both HP and LP devices meet the criteria of the extended ITRS standard. Hence, we selected a doping concentration of $2 \times 10^{13}$ cm$^{-2}$ in the subsequent simulations.

### 3.3 Effects of UL

At the optimized doping concentration of $2 \times 10^{13}$ cm$^{-2}$, we simulated the transfer characteristic curves of *n*-type and *p*-type DG ML MoS$_2$ devices with varying $L_{\text{UL}}$ ($\varepsilon_{\text{s}}$ is fixed at 20), as illustrated in **Figure 2 (a)(d)**. We extracted the on-state current (**Figure 2 (b)(e)**) and SS (**Figure 2 (c)(f)**) from the transfer characteristic curves. The detailed figures of merit are depicted in **Table 1** and **Table 2**.

As depicted in **Figure 2 (a)(d)**, when $L_{\text{UL}}$ is 0.7 Å, due to the significant tunneling effect, the gate exhibits weak control, and turning off the channel becomes challenging. Neither *n*-type nor *p*-type devices' transfer characteristic curves meet the off-state current requirements of 100 nA/μm HP applications and 109 nA/μm for LP applications in the extended ITRS standard. Slightly increasing $L_{UL}$ can significantly improve the gate's control capability, as evidenced by a rapid decrease in SS. When $L_{UL}$ increases from 0.7 Å to 3.2 Å, the SS of the *n*-type and *p*-type devices decrease from 671 mV/dec to 246 mV/dec and from 288 mV/dec to 221 mV/dec, respectively. As illustrated in **Figure 2 (c)(f)**, for both *n*-type and *p*-type devices, the rate at which the SS decreases with the increase of $L_{UL}$ gradually slows down, reaching the lowest SS values of 69 mV/dec and 67 mV/dec at $L_{UL}$ = 25.7 Å, respectively.

The decrease in SS is accompanied by an increase in the on-state current, which plays a crucial role in evaluating the performance of FETs. A high on-state current often corresponds to a high on/off ratio and fast switching speed. At a gate length of 0.34 nm, the extended ITRS standard specifies on-state current requirements of 326 μA/μm for HP applications and 59 μA/μm for LP applications. This implies that devices with an on-state current exceeding these standards can be considered candidates for sub-1-nm transistors. For the *n*-type DG ML MoS$_2$ FETs with a gate length of 0.34 nm, the minimum $L_{UL}$ required to meet the extended ITRS standard for on-state current is 20.7 Å for HP applications and 23.2 Å for LP applications. In contrast, for *p*-type devices, the minimum $L_{UL}$ required to meet the on-state current standards of HP and LP applications is 10.7 Å and 20.7 Å, respectively, which are shorter than the corresponding $L_{UL}$ required for *n*-type devices.

As $L_{UL}$ further increases, the on-state current of *n*-type HP devices reaches its maximum of 409 μA/μm at $L_{UL}$ = 23.2 Å. Subsequently, with further increases in $L_{UL}$ to 25.7 Å, the on-state current slightly decreases to 400 μA/μm. A similar phenomenon is observed in the correlation between the on-state current and $L_{UL}$ for *p*-type HP devices: the on-state current tends to stabilize after $L_{UL}$ reaches 13.2 Å and reaches a maximum of 800 μA/μm when $L_{UL}$ = 20.7 Å. Compared to HP devices, LP devices have lower off-state current standards and are more likely to benefit from the decrease in SS. Hence, within the simulated $L_{UL}$ range (0.7-25.7 Å), we did not observe a reduction in on-state current with the increase of $L_{UL}$. For both *n*-type and *p*-type LP devices, the on-state current reaches its maximum at the longest $L_{UL}$ of 25.7 Å in our simulations, reaching 75 μA/μm and 187 μA/μm, respectively.

## 3.4 Mechanism of UL Effects on Transport

SS is defined as the amount of gate voltage required to increase the current by an order of magnitude in the subthreshold region. SS in **Figure 2 (c)(f)** decreases as $L_{UL}$ increases, and this trend can be explained by a simple square potential barrier model. Here, we use the *n*-type HP device to illustrate the mechanism. In this case, the right electrode serves as the source, and the left electrode serves as the drain. The potential distribution within the device is shown in **Figure S4** (supporting information [47]). According to the Landauer-Büttiker formula [55-57]:

$$I = \frac{2q}{h} \int D(E) T_t(E) [f(E - \mu_S) - f(E - \mu_D)] dE \tag{10}$$

where $I$, $q$, $h$, $D(E)$, $T_t(E)$, $E$, $f$ and $\mu_{S/D}$ are current, carrier charge, Planck constant, mode number, transmission coefficient, energy, Fermi distribution function, and the chemical potential of the source/drain electrode respectively. SS is defined as:

$$SS = \frac{\partial V_g}{\partial (\log_{10} I)} = \frac{\partial q V_g}{\partial \Psi_S} \frac{\partial \Psi_S}{\partial (\log_{10} I) q} = \frac{1}{C} \frac{\partial \Psi_S}{\partial (\log_{10} I) q} \tag{11}$$

In the above statement, $V_g$ and $\Psi_S$ are the gate voltage and the channel surface potential, respectively. $C$ is the body factor, which is dependent on the specific shape of the device and is typically less than 1. $\frac{\partial \Psi_S}{\partial (\log_{10} I) q}$ is the transport factor. The body factor and transport factor together determine SS.

The current consists of two components: thermal current and tunneling current:

$$I = I_{thermal} + I_{tunnel} \tag{12}$$

$$r_{tunnel} = \frac{I_{tunnel}}{I} \tag{13}$$

Among them, $r_{tunnel}$ is the tunneling current rate. Define the SS for thermal current and tunneling current separately as:

$$SS_{thermal} = \frac{\partial V_g}{\partial (\log_{10} I_{thermal})} \tag{14}$$

$$SS_{tunnel} = \frac{\partial V_g}{\partial (\log_{10} I_{tunnel})} \tag{15}$$

It is easy to prove [58]:

$$\frac{1}{SS} = \frac{r_{tunnel}}{SS_{tunnel}} + \frac{1 - r_{tunnel}}{SS_{thermal}} \tag{16}$$

Let's consider the thermal current component first. For a typical FET with metal as the injection material, $D(E)$ is a constant $D_0$, and the transmission coefficient of thermal current is high, close to 1. When the bias voltage is greater than several $k_B T$ (thermal energy unit, where $k_B$ and $T$ are the Boltzmann constant and temperature, respectively), $f(E - \mu_S) - f(E - \mu_D)$ is approximately equal to $f(E - \mu_S)$ in the energy range above the source chemical potential. If we take the source chemical potential $\mu_S$ as the reference point for energy, within an energy range greater than a few $k_B T$, $f(E - \mu_S) \approx \exp(-E/(k_B T))$. In this case, we have:

$$\begin{aligned} I_{\text{thermal}} &= \frac{2q}{h} D_0 \int_{\phi_B} [f(E - \mu_S) - f(E - \mu_D)] dE \\ &\approx \frac{2q}{h} D_0 \int_{\phi_B}^{\infty} \exp(-E/(k_B T)) \, dE \\ &= \frac{2q}{h} D_0 T k_B e^{-\frac{\phi_B}{k_B T}} \end{aligned} \quad (17)$$

Where $\phi_B$ represents the height of the effective potential barrier. In this case:

$$SS_{\text{thermal}} = \frac{\partial V_g}{\partial(log_{10} I_{\text{thermal}})} = \frac{k_B T \ln 10}{q} \frac{\partial V_g}{\partial \Psi_S} \leq \frac{k_B T \ln 10}{q} \quad (18)$$

The term $\frac{k_B T \ln 10}{e}$ is approximately equal to 60 mV/dec at room temperature. 60 mV/dec is the limit of SS due to thermal excitation [59,60].

For the tunneling current component, we use a square potential barrier in place of the actual potential barrier for ease of analysis, as illustrated in **Figure S4** (supporting information [47]). The width and height of the square potential barrier are represented by $\lambda$ and $\phi_B$, respectively. Using longer $L_{\text{UL}}$ implies a wider potential barrier by setting $\lambda = 2L_{\text{UL}} + L_g$. We calculate the transmission coefficient using the Wentzel-Kramer-Brillouin (WKB) approximation [61,62]:

$$T_t(E) = \exp(-\frac{2\lambda}{\hbar}\sqrt{2m_e^*[\phi_B - E]}) \quad (19)$$

Where $\hbar$ and $m_e^*$ represents the reduced Planck constant and the effective mass of the carrier (in this case, the effective mass of the electron $m_e^*$). Substituting equation (19) into equation (10) yields:

$$I_{\text{tunnel}} = \frac{2q}{h} D_0 \int_0^{\phi_B} \exp\left(-\frac{2\lambda}{\hbar}\sqrt{2m_e^*[\phi_B - E]}\right) \frac{1}{\exp(\frac{E}{k_B T})+1} dE \quad (20)$$

In previous studies, the impact of the presence of tunneling current on the gate control has been widely qualitatively discussed [10,58,63,64]. However, quantitative analysis has been lacking. In this section, we utilize a square potential barrier model in conjunction with the WKB approximation to establish a quantitative relationship between the SS, the effective potential barrier height $\phi_B$, the tunneling current rate $r_{tunnel}$, and $L_{UL}$. Equation (20) allows us to obtain the $I_{tunnel}$ vs $\phi_B$ and $SS_{tunnel}$ vs $\phi_B$ curves for various $L_{UL}$, as depicted in **Figure 3**. As $L_{UL}$ increases, the control exerted by $\phi_B$ over the current intensifies, as evidenced by the more rapid decrease of the current with the increasing $\phi_B$ (**Figure 3 (a)**), along with an overall reduction in SS (**Figure 3 (b)**).

We can explore the mechanism of UL effects on transport by comparing the simulation results of quantum transport with the square barrier model. The effective potential barrier height $\phi_B$ of the device can be directly extracted from the projected density of states (PDOS), as exhibited in **Figure 4 (a)(b)** (data are recorded in pdos_currentspectrum_UL.mp4). The current below $\phi_B$ corresponds to tunneling current, indicated in blue in the current spectrum. The current above $\phi_B$ represents thermal transport current, denoted in red in the current spectrum. It can be observed that for the device with $L_{UL} = 20.7$ Å, the current in the off-state is predominantly tunneling current, while in the on-state, it is mostly thermal transport current. The tunneling current rate $r_{tunnel}$ can be calculated as:

$$r_{tunnel} = \frac{I_{tunnel}}{I_{thermal}+I_{tunnel}} \quad (21)$$

We extracted $\phi_B$ of the on and off states for devices with different $L_{UL}$, as displayed in **Figure 4 (c)**, and estimate the average body factor from the change in $\phi_B$ between the on and off states as:

$$\frac{1}{C} = \frac{\partial qV_g}{\partial \Psi_S} \approx \frac{\Delta qV_g}{\Delta \phi_B} \quad (22)$$

It's important to note that due to the difference of 0.54 V in $V_g$ between the on and off state, this is an overall average of the gate voltage's influence on the channel barrier. **Figure 4 (d)** illustrates the increasing $L_{UL}$ leads to a slight reduction in the average body factor.

Using equation (16) and considering the effect of the body factor by equation (11), we can obtain the estimated $SS_{WKB}$ of the square barrier model:

$$SS_{WKB} = \frac{SS_{thermal} SS_{tunnel}}{SS_{thermal} r_{tunnel}+SS_{tunnel}(1-r_{tunnel})} \frac{1}{C} \quad (23)$$

We extract $SS_{tunnel}$ corresponding to $\phi_B$ in the off state from **Figure 3 (b)**, use 60 mV/dec as $SS_{thermal}$, obtain $r_{tunnel}$ from equation (21) (as shown in **Figure 4 (e)**), and estimate $C$ out from equation (22) (as shown in **Figure 4 (d)**). With this information, we can calculate the subthreshold swing $SS_{WKB}$ estimated by the barrier potential model. From the transport curve in **Figure 2 (a)**, we can directly obtain the subthreshold swing $SS_{Quantum}$ from quantum transport simulations. The trends of both $SS_{WKB}$ and $SS_{Quantum}$ with respect to $L_{UL}$ are consistent, and values of $SS_{WKB}$ and $SS_{Quantum}$ in the interval where $L_{UL} > 15.7$ Å almost overlap, as shown in **Figure 4 (f)**. Increasing $L_{UL}$ has two main effects: it reduces the body factor and increases the width of the tunneling barrier. From equation (11), it can be observed that a decrease in the body factor increases the SS. The overall reduction in SS is primarily driven by the stronger effect of the increasing tunneling barrier width, which plays a dominant role.

### 3.5 Compare with Recent Experiments

In their recent experimental advancement, Ren and Tian *et al.* employed ML graphene as the vertical gate, HfO$_2$ as the spacer layer, and ML MoS$_2$ as the channel material to fabricate the FET with the shortest gate length to date, which is just 0.34 nm [17]. Experimental data extracted from this study is presented in **Figure 5**. However, due to the 4 μm channel length and non-Ohmic contacts in this experiment, the devices exhibit poor saturation current and gate control. Experimentally measured saturation current at $V_{ds}$ = 3 V is four orders of magnitude smaller than the saturation current we simulated at $V_{ds}$ = 0.54 V. The experimentally obtained SS values range from 117 to 246 mV/decade, which are approximately 1.7 to 3.6 times higher than the optimized SS of 69 mV/decade in our simulations. The on-state currents extracted from the experimental data are summarized in **Table S2** (supporting information [47]). The on-state currents with $V_{ds}$ = 3 V in the experiments for HP and LP applications are estimated to be $3.1 \times 10^{-1}$ and $2.0 \times 10^{-1}$, respectively. These values fall considerably below the extended ITRS standards and our first-principles simulation results. Our ideal sub-1-nm FET, as simulated, exhibits superior gate control and higher on-state current, showcasing the tremendous potential of sub-1-nm transistors.

### 3.6 Impact of Spacer

In addition to the underlap region, the spacer region is also commonly employed to optimize the transport performance of transistors [53,65]. As an example, we consider the *n*-type HP DG

ML MoS$_2$ FET with a gate length of 0.34 nm and $L_{\text{UL}}$ of 25.7 Å. Different dielectric constants for the spacer regions are configured, and their influence on transport properties is investigated, as illustrated in **Figure 6**. We utilized the dielectric constants of commonly used dielectric materials in experiments for our simulations. The dielectric materials compatible with 2D technology are listed in **Table S3** [66].

The on-state currents of both HP and LP devices in **Figure 6 (b)** increase with the increase in the spacer region's dielectric constant $\varepsilon_s$. The on-state current meets the extended ITRS standard for HP and LP applications when $\varepsilon_s$ is 9 and 15 $\varepsilon_0$, respectively (where $\varepsilon_0$ is the vacuum dielectric constant). This is because the gate control capability improves as $\varepsilon_s$ increases, as indicated by the decrease in SS, as shown in **Figure 6 (c).** We computed the PDOS of the off-state for the device with various $\varepsilon_s$ values, as illustrated in pdos_currentspectrum_spacer.mp4. In contrast to the underlap impact mechanism, which is influenced by both the body factor and the barrier width when changing $L_{\text{UL}}$, it is evident from the PDOS that the variation in barrier width is not substantial when altering $\varepsilon_s$.

We use the HP case as an example to elucidate how the dielectric constant of the spacer region impacts transport properties. We extracted the corresponding effective potential barrier height $\phi_B$ from the PDOS for the on-state and off-state of the devices with various $\varepsilon_s$ values, as depicted in **Figure 7 (a)**. When $\varepsilon_s$ changes, the $\phi_B$ of the off-state remains relatively stable, while the $\phi_B$ of the on-state undergoes more significant variations. This phenomenon arises because that changes in $\varepsilon_s$ alter the gate's control over the channel potential, while the off-state requirement is the same. To quantify this property, we calculated the average body factor between the on-state and off-state using equation (22), as exhibited in **Figure 7 (b)**. It becomes more apparent from **Figure 7 (b)** that increasing $\varepsilon_s$ enhances the body factor.

Similar to the investigation of the underlap mechanism in Section **3.4**, we estimated the SS of the device using equation (23) and the WKB approximation. We compared these estimates with the results obtained from quantum transport, as illustrated in **Figure 7 (c)**. Despite some numerical discrepancies, the WKB model can effectively describe how the SS varies with $\varepsilon_s$ in terms of both trend and magnitude.

## 3.7 Switching Speed and Power Consumption

A transistor's switching process is essentially a charge and discharge operation of its channel. Two crucial parameters for quantifying this process are delay time ($\tau$) and power consumption (PDP), which we calculate using the following formulas [67,68]:

$$\tau = \frac{Q_{on} - Q_{off}}{I_{on}} \qquad (24)$$

$$\text{PDP} = V_{dd}(Q_{on} - Q_{off}) \qquad (25)$$

where $Q_{on}/Q_{off}$, $V_{dd}$, and $I_{on}$ represent the charge in the channel at the on/off state, the driven voltage of 0.54 V, and the on-state current, respectively. $\tau$ and PDP quantitatively represent the time and energy consumption of the switching process, respectively. We calculated these performance metrics for the 0.34 nm gate length DG ML MoS$_2$ transistors, as depicted in **Figure 8**. Detailed parameters are summarized in **Table 1** and **Table 2**.

Overall, there is a general tendency for $\tau$ to decrease with an increase in the underlap length for both *n*-type and *p*-type devices, as shown in **Figure 8 (a)(c)**. This trend is attributed to the increase in on-state current with the increasing $L_{UL}$, leading to faster charging and discharging processes. According to the extended ITRS standard, the delay times for 0.34 nm gate length HP and LP devices should be less than 0.424 and 2.416 ps, respectively. In the case of *n*-type doping, the minimum $L_{UL}$ to meet the extended ITRS standards for HP and LP devices is 13.2 and 18.2 Å, respectively. For *p*-type doping, the minimum $L_{UL}$ for HP and LP devices to comply with the extended ITRS standards is 5.7 and 18.2 Å, respectively. The PDP in **Figure 8 (b)(d)** increases with the increasing $L_{UL}$. The extended ITRS for 0.34 nm gate length devices specify power consumption standards of 0.07 fJ/μm and 0.08 fJ/μm for HP and LP applications, respectively. The power consumption ranges for the *n*-type and *p*-type devices we studied, with $L_{UL}$ ranging from Å 0.7 to 25.7 Å are 0.012-0.048 fJ/μm and 0.008-0.039 fJ/μm, respectively, all of which meet the extended ITRS standards.

Generally, for transistors, lower PDP and faster $\tau$ are preferred. However, these two goals are often conflicting, and it's challenging to achieve both simultaneously. A higher on-state current results in faster switching speed and higher power consumption, as evident in the comparison of HP and LP devices. In comparison to the LP device of the same doping type and $L_{UL}$, the HP device has a higher on-state current, resulting in a smaller $\tau$ but higher PDP. To strike a balance between them, one can utilize the energy-delay product (EDP = PDP × $\tau$) or use the concept of the Pareto front in multi-objective optimization to characterize optimal solutions [69,70].

For the 0.34 nm gate length DG ML MoS$_2$ FETs studied in this paper, we chose parameters that minimize EDP and compared these parameters with the minimum EDP device parameters of FETs made from typical DG ML 2D materials [45,49,58,71-73], as shown in **Figure 9**. It should be noted that, due to the lack of research on sub-1 nm gate length devices, the FET gate lengths in the other studies we have chosen for comparison are all 1 nm. Compared to the DG ML MoS$_2$ FET with a 1 nm gate length, the DG ML MoS$_2$ FETs with a gate length of 0.34 nm in this paper exhibit smaller EDP values for HP *n*-type doping, HP *p*-type doping, LP *n*-type doping, and LP *p*-type doping devices. This demonstrates the potential of sub-1 nm gate length devices in balancing energy consumption and switching speed. The ML MoS$_2$ sub-1 nm gate length devices we studied do not lie on the Pareto front for HP or LP devices shown in **Figure 9**. This is due to the different inherent properties of other typical ML 2D materials, unrelated to device structure optimization. Further research can investigate sub-1 nm gate length devices made from ML silicane [73], zigzag-oriented ML black phosphorus (BP-zig) [58], and zigzag-oriented ML tellurene (Te-zig) [71] that are positioned along the Pareto front.

## 4 Conclusion

In this paper, we employed first-principles quantum transport simulations to investigate the performance limits of sub-1 nm gate length ML MoS$_2$ transistors and used a WKB tunneling model to explain the physical mechanisms of underlap and spacer region optimization on transistor performance. The underlap and spacer regions primarily influence the transport properties of sub-1 nm transistors by adjusting the width and body factor of the potential barriers, respectively. In the case of *n*-type doping, the minimum underlap lengths required to meet the extended ITRS standards for HP and LP applications are 20.7 and 23.2 Å, respectively. The on-state currents of the *n*-type HP and LP devices can reach 409 and 75 μA/μm, respectively. For *p*-type doping, the minimum underlap lengths required to meet the standards of HP and LP devices are 10.7 and 20.7 Å, respectively, which are shorter than those required for the corresponding *n*-type devices. The maximum on-state currents for *p*-type HP and LP applications are 800 and 187 μA/μm, respectively. Compared to the performance limits of ML MoS$_2$ transistors with a 1 nm gate length, sub-1 nm gate length ML MoS$_2$ transistors of the same doping type, both for HP and LP applications, exhibit lower energy-delay products. This suggests that sub-1 nm transistors can better balance energy consumption and switching speed. The quantum transport simulation performance limits we have provided significantly exceed the existing experimental results, indicating that there is substantial room for optimization in sub-1 nm transistors in the experiment. Theoretical and experimental research on sub-1 nm transistors made from other typical 2D semiconductor materials is also highly anticipated.

## Supporting Information

The Supporting Information is available [47]. Two mp4 files: pdos_currentspectrum_UL.mp4: PDOS and current spectra for devices with different $L_{UL}$ in Section 3.4; pdos_currentspectrum_spacer mp4: PDOS and current spectra for devices with different $\varepsilon_s$ in Section 3.6.

## Acknowledgment

This work was supported by the National Key R&D Program of China (Grant No. 2022YFA1203904), the National Natural Science Foundation of China (Grants No. 12274002, No. 11804140, and No. 91964101), the Foundation of He'nan Educational Committee (Grant No. 23A430015), and the High-performance Computing Platform of Peking University. Ying Li is grateful to Dr. Hengwei Luan, Dr. Sixuan Li, and Miss Yuqi Chen for fruitful discussions.

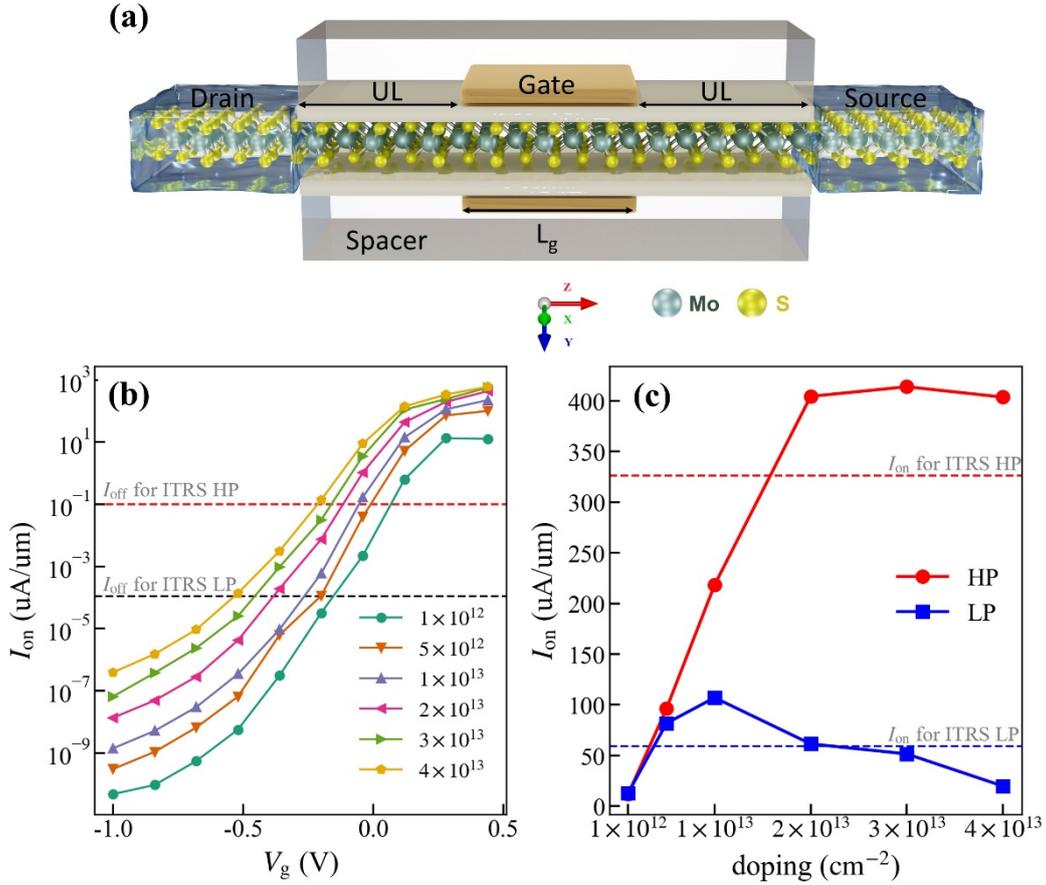

**Figure 1.** Schematic diagram and doping optimization of the DG ML MoS$_2$ FETs with a gate length of 0.34 nm. **(a)** Schematic diagram of the DG ML MoS$_2$ FET. **(b)** Transfer characteristic curves of the DG ML MoS$_2$ FETs at different doping concentrations (1×10$^{12}$ ~ 4×10$^{13}$ cm$^{-2}$). $L_g$ = 0.34 nm, $L_{UL}$ = 23.2 Å, $\varepsilon_s$ = 20. **(c)** The on-state currents are extracted from the transfer characteristic curves. The red and blue dashed lines represent the on-state current standards for high-performance (HP) and low-power (LP) applications specified by the extended ITRS, respectively.

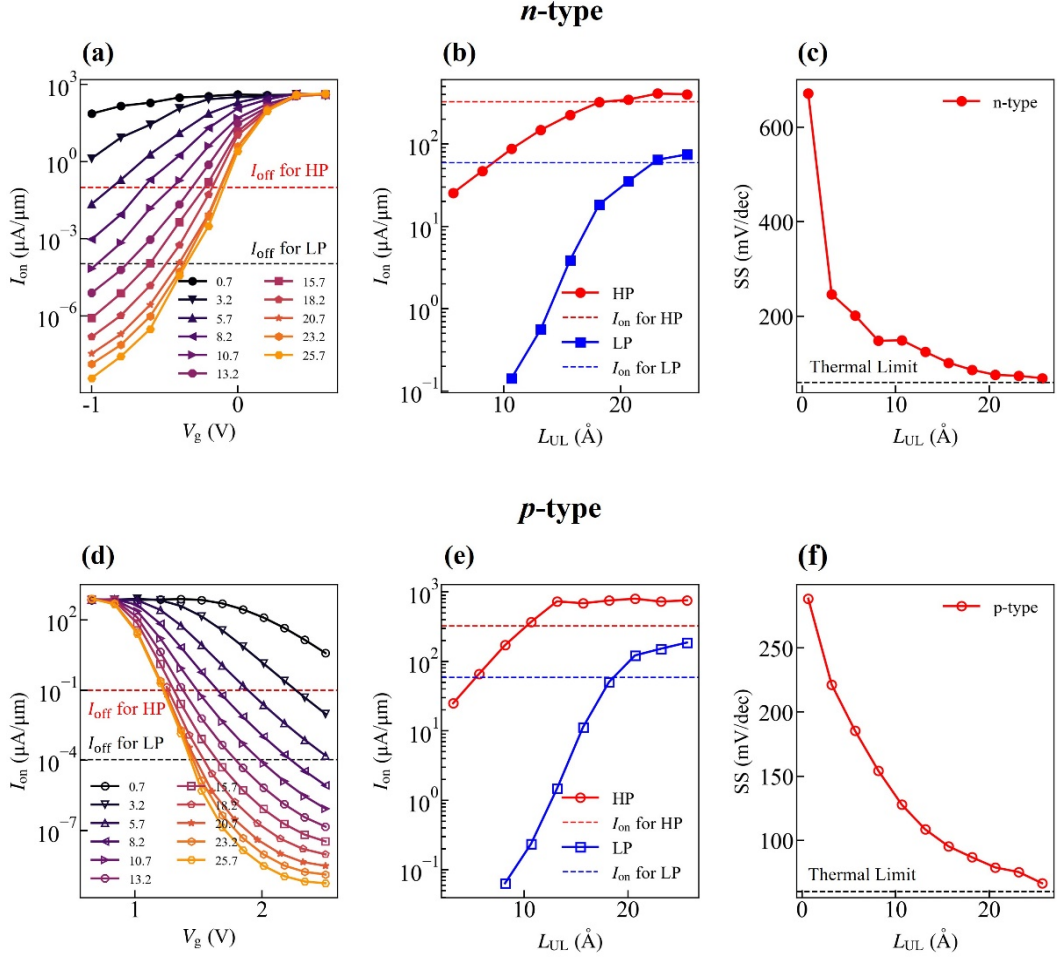

**Figure 2.** Effects of underlaps. The first and second rows contain data for *n*-type devices (solid markers) and *p*-type devices (hollow markers), respectively. The first, second, and third columns represent the transfer characteristic curves, on-state currents, and subthreshold swing (SS), respectively. The extended ITRS standard and the thermal transport limit of the SS are indicated by dashed lines. Different lengths of underlaps $L_{\text{UL}}$ in (a) and (d) are marked with different markers and colors.

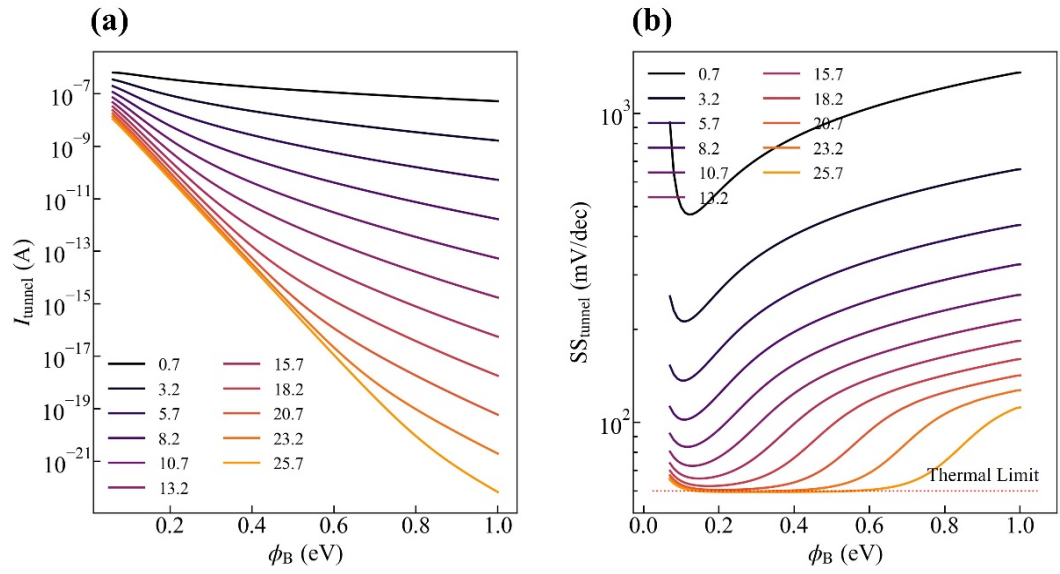

**Figure 3.** Tunneling current and tunneling subthreshold swing (SS) as functions of the effective potential barrier $\phi_B$ under different $L_{UL}$ in the square potential barrier model. The width of the barrier is determined by $\lambda = 2L_{UL} + L_g$. $L_{UL}$ is denoted in the labels.

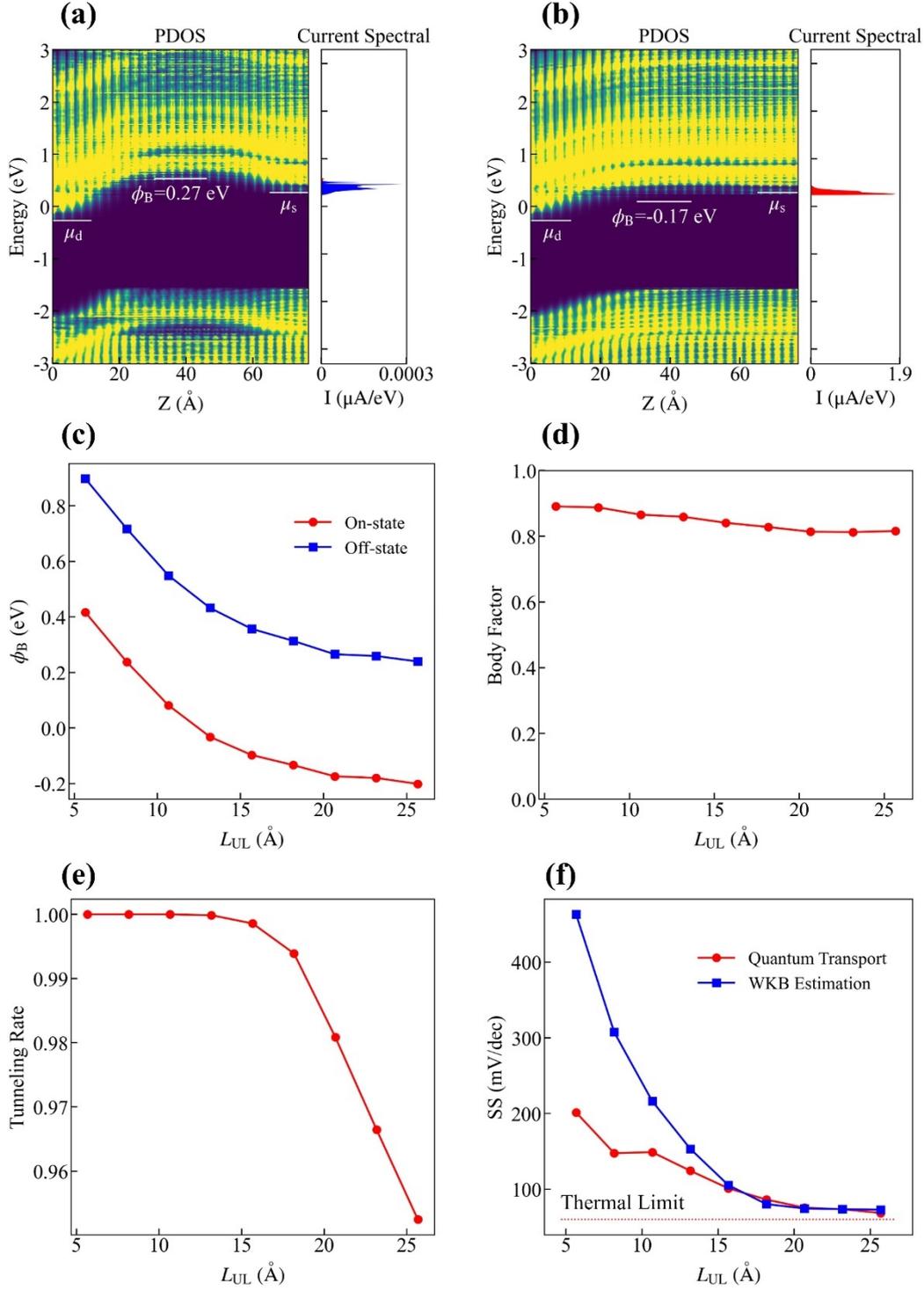

**Figure 4.** Mechanism of UL effects on transport, illustrated with *n*-type HP devices. Schematic diagrams of the projected density of states (PDOS) and current spectrum for the off-state **(a)** and on-state **(b)** of the device with $L_{UL}$ = 20.7 Å. **(c)** The effective barrier $\phi_B$ of the device in the on and off states varies with $L_{UL}$. The average body factor in the switching voltage range **(d)** and the tunneling current ratio in the off-state **(e)** vary with $L_{UL}$. **(f)** Comparison of subthreshold swing (SS) obtained from quantum transport and WKB approximation calculations.

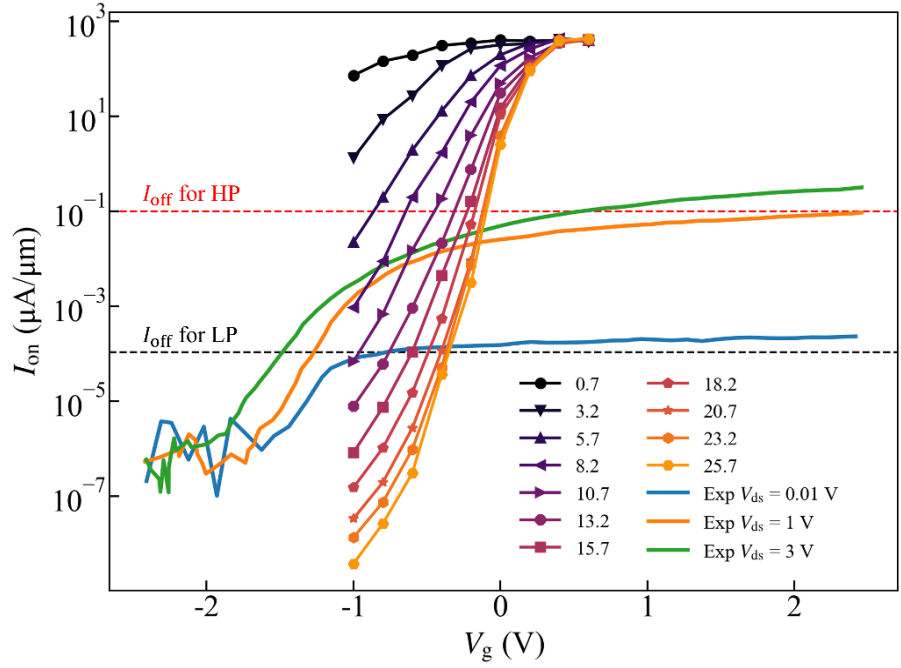

**Figure 5.** Comparison of quantum transport simulation with recent experiments. The green, orange, and blue curves, unmarked, represent $V_{ds}$ = 0.01 V, 1 V, and 3 V from experiments [17], respectively. Our quantum transport simulation results with different $L_{UL}$ values are indicated by various markers and colors.

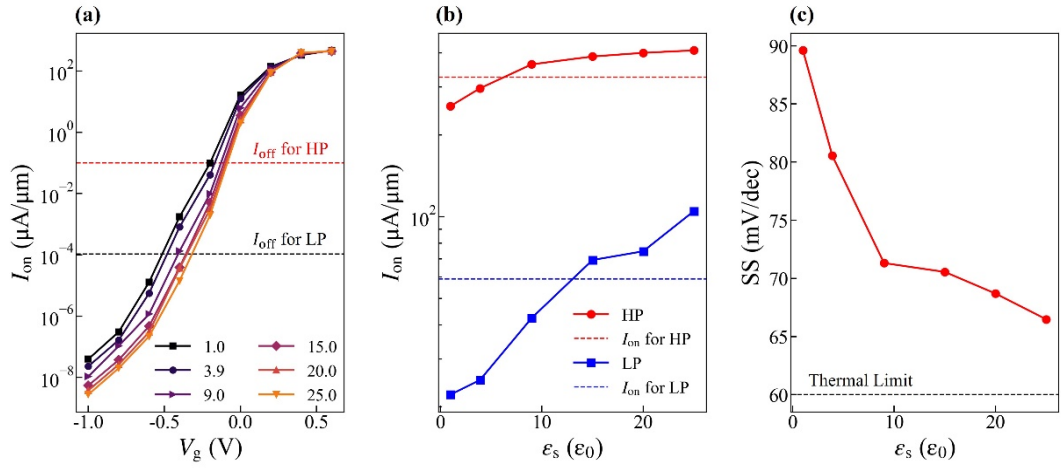

**Figure 6.** Impact of spacer region dielectric constant on transport properties. Illustrated with an *n*-type 0.34 nm gate length DG ML MoS$_2$ FET with a $L_{UL}$ of 25.7 Å as an example. **(a)** Transfer characteristic curves of devices with spacer regions featuring varying dielectric constants ($\varepsilon_s$). **(b)** On-state currents for HP and LP applications extracted from **(a)**. **(c)** Subthreshold swing (SS) extracted from **(a)**.

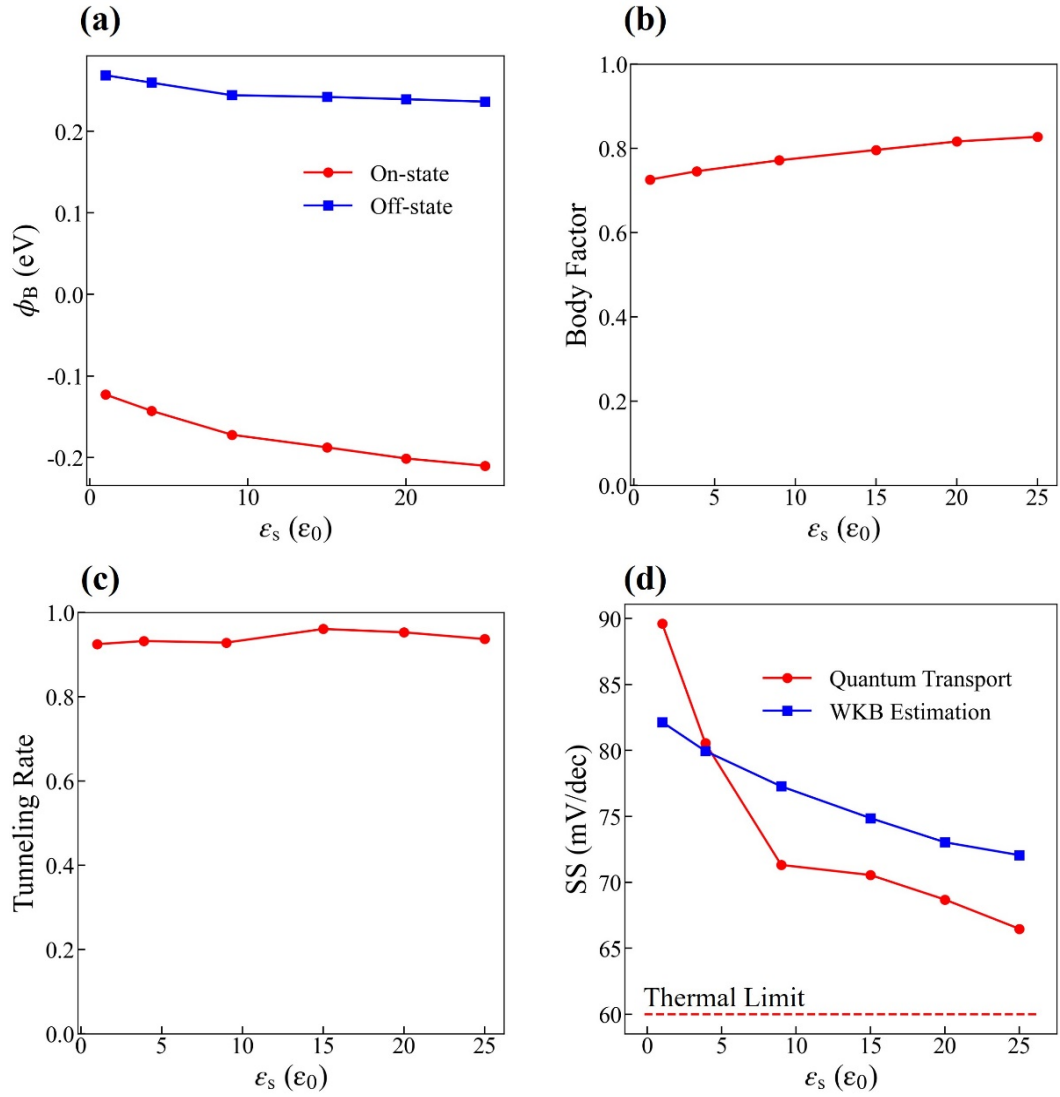

**Figure 7.** Mechanism of dielectric constant ($\varepsilon_s$) in the spacer region: taking the *n*-type HP DG ML MoS$_2$ FETs with $L_{UL}$ = 25.7 Å as an example. **(a)** Effective potential barrier height in the on-states and off-states. **(b)** Average body factor across the switching voltage range. **(c)** Tunneling current ratio in the off-state. **(d)** Comparison of subthreshold swing (SS) obtained from quantum transport and WKB approximation calculations.

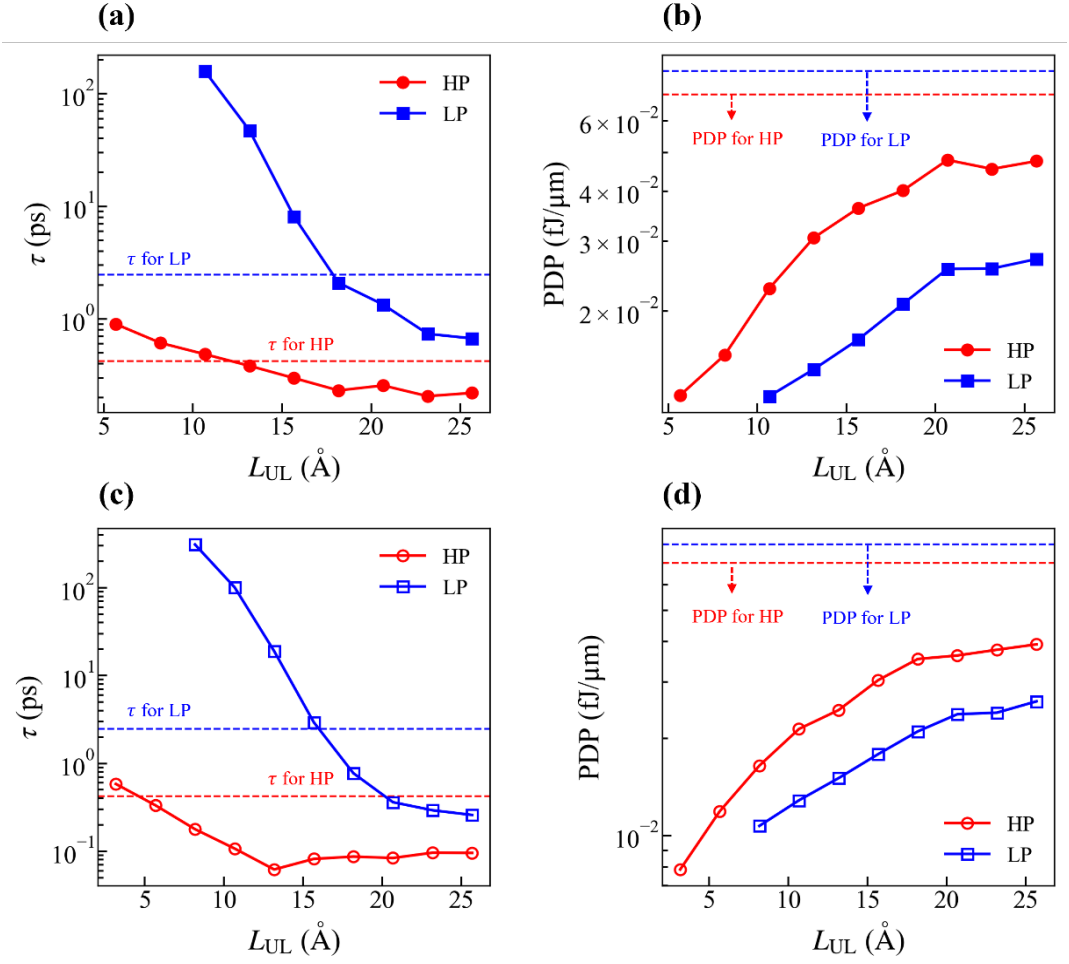

**Figure 8.** Switching speed and power consumption. The first row contains data for *n*-type doping (solid icons), and the second row contains data for *p*-type doping (hollow icons). The first and second columns respectively represent the delay time $\tau$ and power consumption PDP. The expanded ITRS standard values are marked with dashed lines in the figure.

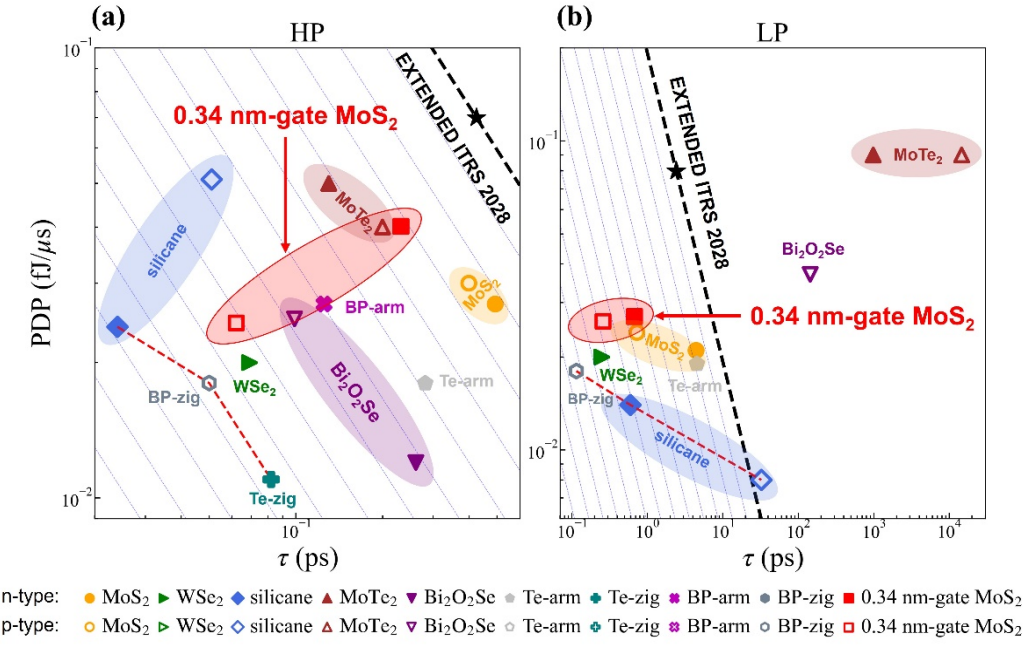

**Figure 9.** PDP vs $\tau$. Comparison between DG ML MoS$_2$ FETs with a 0.34 nm gate length and typical DG ML 2D material FETs with a 1 nm gate length [45,49,58,71-73]. The red dashed lines represent the Pareto fronts. Filled and hollow markers correspond to *n*-type and *p*-type doping, respectively. Abbreviations are used: tellurene (Te), black photosphere (BP), armchair direction (arm), zigzag direction (zig).

**Table 1.** Comparison of performance parameters for *n*-type and *p*-type 0.34 nm gate length DG ML MoS$_2$ FETs and extended ITRS standards for HP applications. $L_g$: gate length; $L_{UL}$: Underlap length; SS: Subthreshold swing; $I_{on}$: on-state current; $I_{on}/I_{off}$: on-off ratio; $\tau$: delay time; PDP: power dissipation.

| $L_g$ (Å) | $L_{UL}$ (Å) | SS (mV/dec) | | $I_{on}$ (µA/µm) | | $I_{on}/I_{off}$ | | $\tau$ (ps) | | PDP (fJ/µm) | |
|---|---|---|---|---|---|---|---|---|---|---|---|
| | | *n* type | *p* type | *n* type | *p* type | *n* type | *p* type | *n* type | *p* type | *n* type | *p* type |
| | 0.7 | 671 | 288 | - | - | - | - | - | - | - | - |
| | 3.2 | 246 | 221 | - | 25 | - | 2.49×10² | - | 0.584 | - | 0.008 |
| | 5.7 | 201 | 185 | 25 | 66 | 2.54×10² | 6.61×10² | 0.899 | 0.334 | 0.012 | 0.012 |
| | 8.2 | 148 | 154 | 47 | 171 | 4.70×10² | 1.71×10³ | 0.612 | 0.178 | 0.016 | 0.016 |
| | 10.7 | 149 | 128 | 87 | 369 | 8.70×10² | 3.69×10³ | 0.486 | 0.107 | 0.023 | 0.021 |
| 3.4 | 13.2 | 124 | 109 | 148 | 732 | 1.48×10³ | 7.32×10³ | 0.383 | 0.062 | 0.031 | 0.025 |
| | 15.7 | 101 | 95 | 225 | 687 | 2.25×10³ | 6.87×10³ | 0.298 | 0.082 | 0.036 | 0.030 |
| | 18.2 | 86 | 87 | 321 | 752 | 3.21×10³ | 7.52×10³ | 0.232 | 0.087 | 0.040 | 0.035 |
| | 20.7 | 76 | 79 | 346 | 800 | 3.46×10³ | 8.00×10³ | 0.256 | 0.084 | 0.048 | 0.036 |
| | 23.2 | 74 | 75 | 409 | 725 | 4.09×10³ | 7.25×10³ | 0.206 | 0.096 | 0.045 | 0.038 |
| | 25.7 | 69 | 67 | 400 | 758 | 4.00×10³ | 7.58×10³ | 0.220 | 0.096 | 0.048 | 0.039 |
| 3.4 (ITRS) | - | - | | 326 | | 3.26×10³ | | 0.424 | | 0.07 | |

**Table 2.** Comparison of performance parameters for *n*-type and *p*-type 0.34 nm gate length DG ML MoS$_2$ FETs and extended ITRS standards for LP applications. $L_g$: gate length; $L_{UL}$: Underlap length; SS: Subthreshold swing; $I_{on}$: on-state current; $I_{on}/I_{off}$: on-off ratio; $\tau$: delay time; PDP: power dissipation.

| $L_g$ (Å) | $L_{UL}$ (Å) | SS (mV/dec) | | $I_{on}$ (μA/μm) | | $I_{on}/I_{off}$ | | $\tau$ (ps) | | PDP (fJ/μm) | |
|---|---|---|---|---|---|---|---|---|---|---|---|
| | | *n* type | *p* type | *n* type | *p* type | *n* type | *p* type | *n* type | *p* type | *n* type | *p* type |
| 3.4 | 0.7 | 671 | 288 | - | - | - | - | - | - | - | - |
| | 3.2 | 246 | 221 | - | - | - | - | - | - | - | - |
| | 5.7 | 201 | 185 | - | - | - | - | - | - | - | - |
| | 8.2 | 148 | 154 | - | 0.1 | - | 5.9×10$^2$ | - | 310.4 | - | 0.011 |
| | 10.7 | 149 | 128 | 0.1 | 0.2 | 1.3×10$^3$ | 2.2×10$^3$ | 158.3 | 101.1 | 0.012 | 0.013 |
| | 13.2 | 124 | 109 | 0.6 | 1 | 5.2×10$^3$ | 1.4×10$^4$ | 47.21 | 18.96 | 0.014 | 0.015 |
| | 15.7 | 101 | 95 | 4 | 11 | 3.6×10$^4$ | 1.0×10$^5$ | 8.127 | 2.942 | 0.017 | 0.018 |
| | 18.2 | 86 | 87 | 18 | 50 | 1.7×10$^5$ | 4.6×10$^5$ | 2.094 | 0.774 | 0.021 | 0.021 |
| | 20.7 | 76 | 79 | 35 | 122 | 3.2×10$^5$ | 1.1×10$^6$ | 1.341 | 0.362 | 0.026 | 0.024 |
| | 23.2 | 74 | 75 | 64 | 152 | 5.9×10$^5$ | 1.4×10$^6$ | 0.737 | 0.293 | 0.026 | 0.024 |
| | 25.7 | 69 | 67 | 75 | 187 | 6.8×10$^5$ | 1.7×10$^6$ | 0.671 | 0.259 | 0.027 | 0.026 |
| 3.4 (ITRS) | - | - | | 59 | | 5.41×10$^5$ | | 2.416 | | 0.08 | |